\documentclass[prl,twocolumn,amsmath,amssymb,floatfix]{revtex4}
\usepackage{graphicx}
\usepackage{bm}

\newcommand{\beq}{\begin{eqnarray}}
\newcommand{\eeq}{\end{eqnarray}}

\def \be{\begin{equation}}
\def \ee{\end{equation}}
\def \ba{\begin{array}}
\def \ea{\end{array}}
\def \bea{\begin{eqnarray}}
\def \eea{\end{eqnarray}}
\def \nn{\nonumber}
\def \l{\left}
\def \r{\right}
\def \half{{1\over 2}}
\def \etal{{\it {et al}}}

\def \W{{\Omega}}
\def \e{{\epsilon}}

\def \a{{\alpha}}

\def \b{{\beta}}

\def \D{{\Delta}}
\def \d{{\delta}}

\def \f{{\varphi}}

\def \G{{\Gamma}}
\def \z{{\zeta}}

\def \av#1{{\langle#1\rangle}}

\begin{document}

\title{Phase transition of one dimensional bosons with strong disorder}

\author{Ehud Altman$^1$, Yariv Kafri$^1$, Anatoli Polkovnikov$^1$,
Gil Refael$^2$\\
{$^1$\small \em Department of Physics, Harvard University,
Cambridge, MA
  02138}\\
{$^2$\small \em Kavli Institute of Theoretical Physics, University
of California, Santa
  Barbara, CA 93106}}

\begin{abstract}
We study one-dimensional disordered bosons at large commensurate
filling. Using a real-space renormalization group approach, we
find a new {\it random} fixed point which controls a phase
transition from a superfluid to an incompressible Mott-glass. The
transition can be tuned by changing the disorder distribution even
with vanishing interactions. We derive the properties of the
transition, which suggest that it is in the Kosterlitz-Thouless
universality class, and discuss its physical origin.
\end{abstract}

\maketitle

Models of interacting bosons subject to quenched disorder are
important for understanding many real systems and present a
considerable theoretical challenge. Bosons in a disordered
potential, for instance, were studied extensively
~\cite{gs,ma,fwgf,singh,monien,phillips,Moore-BrayAli} in relation
to granular superconductors, Josephson-junction arrays, and $^4He$
on porous media. Recently, there has been interest in this problem
in the context of experiments with ultra cold
atoms~\cite{zoller,charles}; These may provide a controllable
environment for studying quenched disorder and interactions. In
this letter we take a new perspective, and utilize the real-space
renormalization group (RSRG) method
\cite{dasgupta,fisher,fisher-ising}. Using this method it was
shown that random spin chains often exhibit a universal behavior
independent of the disorder realization. Using RSRG we describe
the properties of a superfluid-insulator transition of the {\it
random} $O(2)$ quantum rotor model, and find that it exhibits a
similar degree of universality.

A standard model of granular superconductors and Josephson arrays
in 1-d is the $O(2)$ quantum rotor Hamiltonian: \be H=\half\sum_j
U_j \left(-i\frac{\partial}{\partial\f_j}\right)^2 - \sum_j J_j
\cos(\f_{j+1}-\f_j). \label{model} \ee where $U_i$ is the grain
charging energy, and $J_i$ is the Josephson coupling between
grains. For uniform couplings, the zero-temperature quantum
partition function of the chain can be mapped to that of the two
dimensional classical x-y model with unit coupling and effective
temperature $T^{eff}=\sqrt{U/J}$ \cite{girvin}. It follows that
the pure system displays a Superfluid-Insulator transition of the
Kosterlitz-Thouless (KT) universality class at a critical
interaction strength $\sqrt{U/J}\sim 1$.

The effect of disorder on (\ref{model}) is important for
understanding real systems. In systems such as granular
superconductors, fluctuations in grain size, and distance between
grains are the main source of randomness, which is exhibited in
$J$ and $U$.  The analysis of the random model is often carried
out using the uniform model because randomness in $U$ and $J$ is
perturbatively irrelevant. But the irrelevance of weak disorder
does not rule out fixed points, and hence possible new phase
transitions, at finite disorder. We add that diagonal (potential)
disorder can become relevant above a critical interaction
strength~\cite{gs}. Here assume that our model is strictly
particle-hole symmetric, and omit diagonal disorder form
(\ref{model}).

In this letter we identify a new quantum phase transition from a
disordered superfluid to an incompressible gapless phase
(Mott-glass) using the RSRG method. Remarkably, the fixed point
that controls the transition corresponds to a classical model
($U_i=0$). It can thus be tuned by varying the disorder in $J$
even at arbitrarily small charging energy $U$, in sharp contrast
with the uniform case. We study the universal properties of the
transition using the RG and show that it is KT like with a
dynamical exponent $z=1$, as in the uniform case. In addition to
the RG treatment we present a simple physical picture of the
emerging phases and the phase transition.

\begin{figure}[ht]
\includegraphics[width=7.5cm]{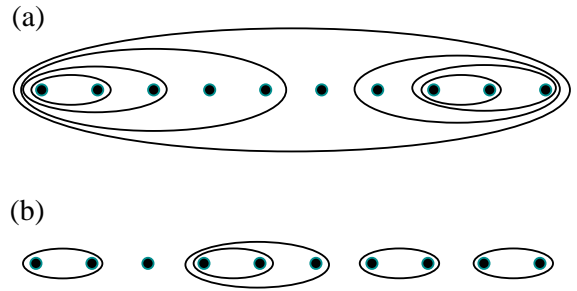}
\caption{Possible fates of the chain after repeated application of
the RG decimation rules. (a) In the superfluid phase sites are
joined into ever growing clusters. (b) Insulating phase, where
clusters become effectively disconnected at low energy scales.}
\label{fig:scheme}
\end{figure}

The Hamiltonian (\ref{model}) is equivalent to the Bose-Hubbard
model at large commensurate filling~\cite{Sachdev_book}, but it
differs in some respects from the traditional dirty boson
problem~\cite{gs,fwgf,monien}. First, the Hamiltonian
(\ref{model}) obeys a local particle hole symmetry. Thus, it
describes only the Superfluid-Insulator transition at commensurate
filling. In addition, it does not contain the diagonal disorder $i
v_j ~\partial/\partial\f_j$. The significance of this difference
was illuminated by recent Monte-Carlo simulations of large two
dimensional systems, which considered both types of
disorder~\cite{prokof}. While the diagonal disorder is expected to
ultimately dominate, its effects may be unobservable at practical
temperatures or length scales.

Let us begin our analysis by describing the RSRG method as applied
to the random-boson problem. In the spirit of
Ref.~[\onlinecite{fisher-ising}], we define the cutoff energy
scale $\Omega=max_i\l\{U_i,\,J_i\r\}$. At each RG step, the chain
element
 with largest energy is decimated, renormalizing nearby
interaction constants. Two types of steps are possible. If
$\W=U_i$ for some site $i$, this site is assumed frozen in its
lowest energy charge state. A connection between nearest neighbors
of the site is generated perturbatively: \be
\tilde{J}_{i-1,i+1}=J_{i-1}J_{i}/\W. \label{jrecurs} \ee In the
alternative case, when $\W=J_i$ the relative phase $\f_{i+1}-\f_i$
is assumed frozen in its ground state and the two sites are
renormalized into a single one. The charging energy of the new
site corresponds to additive recursion relations in terms of
capacitances $C_i\equiv \W/U_i$: \be \tilde{C}_i=C_i+C_{i+1},
\label{crecurs} \ee while the hopping connecting the new site with
its neighbors is unchanged.

Repeated decimations gradually reduce the energy scale from its
initial value $\W_0$ to a lower energy scale $\W$. Depending on
the initial distributions of the couplings, two scenarios emerge.
In the first, sites are joined together into ever growing
superfluid clusters (see Fig. \ref{fig:scheme}(a)) leading to a
superfluid phase. In the second, a growing number of sites are
eliminated and form disconnected clusters leading to an insulating
phase (Fig. \ref{fig:scheme}(b)).

It is convenient to describe the progression of the RG
transformations and the cutoff energy scale with the parameter
$\G\equiv\log(\W_0/\W)$. In addition, we define the dimensionless
coupling $\z_i=C_i-1$ and $\b_i=\log(\W/J_i)$, characterized by
probability distributions $f(\z,\G)$ and $g(\b,\G)$~\cite{note}.
Their flow with the decreasing energy scale is given by the master
equations:
\begin{widetext}
\beq &&{\partial f(\z,\G)\over \partial \G}= (1+\z){\partial
f(\z,\G)\over\partial \z}+g_0(\G)\int d\z_1 d\z_2 f(\z_1,\G)
f(\z_2,\G)\delta(\z_1+\z_2+1-\z)+f(\z,\G)\l(f_0(\G)+1-g_0(\G)\r),
\nn\\
&&{\partial g(\beta,\G)\over \partial \G}= {\partial
g(\beta,\G)\over\partial \beta}+f_0(\G)\int d\beta_1 d\beta_2
g(\beta_1,\G) g(\beta_2,\G)\delta(\beta_1+\beta_2-\beta)
+g(\beta,\G)\l(g_0(\G)-f_0(\G)\r), \label{master} \eeq
\end{widetext}
where $g_0(\G)\equiv g(0,\G)$ and $f_0(\G)\equiv f(0,\G)$. The
first term in each equation describes the flow of the distribution
due to a redefinition of the cutoff following an elimination of
large couplings. The second term implements the recursion
relations, taking account of the renormalized couplings. The last
term removes the couplings neighboring the decimated ones and
takes care of normalization of the distributions.

When either typical $\z\gg 1$ or $\b\gg 1$ one can use
$\delta(\z_1+\z_2+1-\z)\approx\delta(\z_1+\z_2-\z)$ in
(\ref{master}). Then the master equations are solved by the
ansatz: \be \ba{cc} f(\z,\G)=f_0(\G)\mathrm e^{-\z f_0(\G)},&
g(\beta,\G)=g_0(\G)\mathrm e^{- \beta g_0(\G)}, \label{ansatz} \ea
\ee where $f_0$ and $g_0$ obey: \be \ba{cc} {df_0(\G)\over
d\G}=f_0(\G)-f_0(\G)g_0(\G), & {dg_0(\G)\over
d\G}=-g_0(\G)f_0(\G). \label{flow2} \ea \ee Thus, \be
f_0(\G)=g_0(\G)-\ln g_0(\G)+A. \label{eq3} \ee
\begin{figure}[ht]
\includegraphics[width=7.5cm,height=5cm]{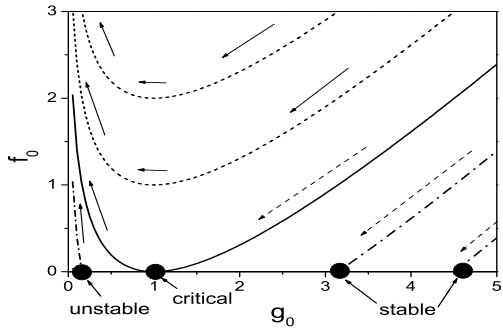}
\caption{Phase space flow of the functions $f_0(\G)$ and
$g_0(\G)$. Different lines correspond to different values of $A$
in (\ref{eq3}). $A>-1$ (dashed lines) result in flow to the
insulating phase. $A<-1$ (dash-dotted lines) flows either to the
insulating phase for small $g_0$, or to the disordered superfluid
phase for large $g_0$. The solid line $A=-1$ is the separatrix.
The line $f_0\equiv 0$ is a line of fixed points which are stable
for $g_0>1$ and unstable for $g_0<1$.} \label{fig:flow}
\end{figure}

It is interesting to note that Eqs. (\ref{flow2}) acquire
precisely the form of the KT flow equations~\cite{jose} when
written in terms of $y=\sqrt{f_0}$ and $x=g_0$. The integration
constant $A$ controls the flow as depicted in Fig.~\ref{fig:flow}.
Distributions with $A>-1$ flow to $f_0\to \infty$ and $g_0\to 0$.
This corresponds to vanishing Josephson coupling and $U\to\Omega$,
namely an insulating phase. For $A< -1$ and $g_0>1$ the flow
approaches a line of fixed points with a non universal
$g_0(\infty)=1+\alpha$, $\a> 0$ and $f_0(\infty)=0$. This
corresponds to an array of Josephson junctions with no charging
energy and a power-law distribution of couplings
$p(J)\propto(J/\W)^\alpha$. The critical flow occurs when $A=-1$.
It terminates at the unstable fixed point $g^*=g_0(\infty)=1$
($\a=0$) and $f^*=f_0(\infty)=0$. Note that even a system with
vanishingly small charging energy and random tunneling can be
tuned through the critical point by changing the disorder
distribution. When $\a<0$ the system flows to the insulating
phase.

We now use Eqs. (\ref{flow2}) together with the asymptotic forms
of $g_0(\G),\,f_0(\G)$ to show that the transition has KT-like
universal properties and dynamical exponent $z=1$. First, we
establish a connection between energy and length scales. As the RG
flow proceeds, more sites are either eliminated or joined into
larger superfluid clusters. One can see that the total number of
sites at scale $\Gamma$ decreases as \be N(\G)=N_0
e^{-\int_0^\G(f_0+g_0)d\G'}. \label{Ntau} \ee
The length scale associated with $\G$ is the size of superfluid
clusters (or bonds between them). In units of the original lattice
constant it is given by $l(\G)$=$
N_0/N(\G)=\exp(\int_0^\G(f_0+g_0)d\G')$. Solving the flow Eqs.
(\ref{flow2}) near the fixed point, $g_0^*=1,~f_0^*=0$ we identify
typical time and length scales, which characterize the critical
flow: \be \ba{cc} \tau_c\propto e^{-1/\sqrt{\e}}, &  \xi_c\propto
e^{1/\sqrt{\e}}, \ea \ee where $\epsilon=1+A>0$ is the tuning
parameter. The exponential divergence toward the critical point
suggests a transition of the KT universality class with a
dynamical exponent $z=1$.

Using the universal distributions we found above we can calculate
several observables. For example, to find the compressibility, we
apply the RG to a finite chain down to the scale $\G_1$, where
only a single site is left ($N(\Gamma_1)=1$). The compressibility
extracted from the renormalized Hamiltonian of the remaining site
$H_1=U n^2$ then reads: \be \kappa=\langle\frac{1}{N_0
U}\rangle\approx \frac{1}{N_0\W(\G_1)}\int_0^\infty
(\z+1)f_0(\G_1)e^{-f_0(\G_1)\z}d\z. \ee Using Eqs. (\ref{Ntau}),
 and (\ref{flow2}) to derive the asymptotic forms of
$f_0(\G)$ and $g_0(\G)$, we find in the insulating regime,
$\kappa_{in}\propto (\ln N_0)/N_0$, which vanishes in the
thermodynamic limit. Similarly, in the SF side
$\kappa_{sf}=k/\Omega_0$, where $k$ is a positive non-universal
constant, that depends on initial distributions. It follows that
$\kappa$ is discontinuous at the critical point. The gap in the SF
phase obeys $\D=\W(\G_1)f_0(\G_1)(-\ln{f_0(\G_1)})\sim
\sqrt{2\epsilon} {\ln N_0\over N_0}$. In the insulating phase, the
gap is $\D\sim\W(\G_1)\sim 1/\ln N_0$. The gap vanishes in this
case because of rare large clusters that have arbitrarily small
charging energy. Thus, we identify the insulating phase with a
Mott glass.

\begin{figure}[h]
\includegraphics[width=7.5cm]{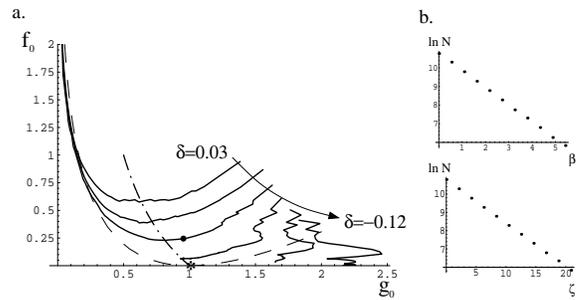}
\caption{Numerical RG flows for systems with initial box
distributions. (a) Distribution of $U$ is centered around
$0.7+\d/2$ and that of $J$ around $0.7-\d/2$ with
$\d=0.03,\,0,\,-0.03,\,-0.06,\,-0.09,\,-0.12$ . The width of both
distributions is always taken to be $1$. $f_0$ and $g_0$ are
defined as the inverse of the numerical average of $\z$ and $\b$
respectively. The dashed line marks the superfluid-insulator
separatrix according to Eq. (\ref{eq3}) with $A=0$.  (b) A semilog
plot of the distributions of the couplings $\z=\W/U-1$ (top) and
$\beta=\ln \W/J$ (bottom) at a typical point in the flow not far
from the critical point (marked by a black dot in (a)).}
\label{fig:num-flow}
\end{figure}

We have implemented the RG scheme numerically to verify that the
solution (\ref{ansatz}) is an attractor of the flow for generic
initial disorder. We started from chains with $2.5\times 10^6$
sites, and different forms of initial disorder distributions. The
general features of the flow diagram seem insensitive to the type
of distributions used. A specific example calculated from various
box distributions of $J$ and $U$ is given in Fig.
\ref{fig:num-flow}. The result clearly shows that the
distributions converge to the solutions (\ref{ansatz}),
demonstrating that they are stable attractors of the RG flow. The
positions of the minima marked by the dashed line in Fig.
\ref{fig:num-flow} agree with those predicted by the first
correction to (\ref{flow2}) from expansion of the $\d$-function in
Eq.~(\ref{master}) to lowest order in $1/\zeta$.

The validity of the recursion relations (\ref{jrecurs}) and
(\ref{crecurs}) relies on the distribution of $J$'s being wide in
the sense that $\av{J}<<\W$. Clearly this does not hold at the
critical point where $\av{J}=\W/2$. While the relation
(\ref{jrecurs}) can be controlled by the smallness of $\av{U}$
near the critical point, corrections to (\ref{crecurs}) produce
next nearest neighbor Josephson couplings of order $\av{J}^2/\W$.
To justify the RG scheme used here, we need to show that the
longer range terms are irrelevant. The corresponding analysis is
rather complicated and we defer it to a future work. Instead, we
rederive the phase transition and explain its nature using an
alternative physical argument.

To understand the origin of the fixed point distributions we
employ two simple toy models (see also Ref.
\onlinecite{Moore-BrayAli}). First, consider the insulating fixed
point. We assume that the Josephson couplings take the values
$J_i=1$ with probability $q$, and $J_i=0$ with probability $1-q$.
We also assume that the charging $U_i=u \ll 1$ is uniform. Such a
chain consists of disconnected clusters and can describe only the
insulating phase. The cluster sizes are distributed according to
$P(N)=q^N$, and the effective interaction is given by $U(N)=u/N$.
From this we immediately see that $f(\zeta)$ is indeed given by
(\ref{ansatz}) with $f_0=|\ln q|$. If we now start eliminating
sites with large $U$ from the chain, we will exactly reproduce the
flow equation in (\ref{flow2}). Clearly, the insulating phase is
stable towards weak Josephson couplings between the clusters.
Therefore the disconnected chain used in this toy model is fairly
generic once we deal with an insulator.

\begin{figure}[ht]
\includegraphics[width=7.5cm]{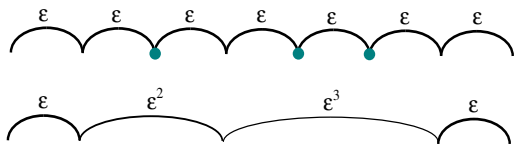}
\caption{Schematic representation of the second toy model. The
circles on the top graph denote sites with a large charging energy
$U=1$. After eliminating these sites we arrive to a classical
chain with disordered couplings between the neighboring sites
(bottom plot).} \label{fig:scheme2}
\end{figure}

We can describe the superfluid fixed line by a chain with uniform
Josephson couplings $J_i=\epsilon<<1$ and charging energies with a
binary distribution: $U_i=1$ with probability $q$, and $U_i=0$
with probability $1-q$. We use perturbation theory to eliminate
the sites with $U_i=1$. Two remaining sites, separated by a block
of $N$ eliminated sites are effectively connected by $J=\e^N$ (see
Fig.~\ref{fig:scheme2}).
In addition, the size distribution of blocks with $U_i=1$ is
$P(N)\sim q^N$. The system is thus described by an effective {\em
classical} chain of Josephson junctions (no charging energy) with
a power-law coupling distribution: \be
P(J)=P(N)\frac{dN}{dJ}=(\a+1)(J/\e)^\a, \label{powerlaw} \ee where
$\a= \ln q/\ln \e -1$.

The toy model suggests that the classical chain with $\alpha>0$ is
also a stable fixed line due to the vanishing probability of
finding a $J=0$ on the chain. A small $U$ perturbation, however,
destabilizes the line for $\alpha<0$. Physically, this is due to
vanishing thermodynamic stiffness; we prove this in the following.
The stiffness $\rho(L)=L(E(\theta)-E(0))/\theta^2$ of a given
realization of a chain of length $L$ is given by
$\rho(L)=L/(\sum_i 1/J_i)$. In the limit $L\to \infty$, the
average over the chain appearing in the denominator can be
replaced by an average over the distribution yielding
$\rho(\infty)=max\{0,\a/(\a+1)\}$. Clearly, a state with zero
stiffness is expected to be unstable to adding a small charging
energy $\Delta_0$. To see this more explicitly, consider
perturbing the system with a small uniform $U_i=\D_0$. We choose
an energy scale $\D_0<\W\ll\W_0=\e$. The weak bonds with $J<\W$
are separated by clusters of average size $L=(\W_0/\W)^{\alpha+1}$
consisting only of stronger bonds. Since each cluster is
superfluid by the choice of parameter, the charging energy of a
cluster of size $l$ is given by $\D(l)\sim \D_0/l$. Because
clusters are independent from each other, the distribution of
their sizes is simply $P(l)=L^{-1}\exp(-l/L)$, the average
charging energy on the coarse grained lattice is $\D\sim\D_0 \ln
L/L$. Putting everything together we find: \be
\frac{\D}{\W}\sim\left(\frac{\D_0}{\W_0}\right)
\left(\frac{\W}{\W_0}\right)^\a \ln\left(\frac{\W_0}{\W}\right)
=\left(\frac{\D_0}{\W_0}\right)\G e^{-\a\G}. \ee Charging energy
is thus relevant for $\a\le 0$, and irrelevant for $\a>0$, in
agreement with the RG flows in Fig. \ref{fig:flow}.

The flow towards an insulator for $\a<0$ implies that the
superfluid stiffness is suppressed exponentially with the system
size. But when $\a=0^+$, $\rho(L)\sim 1/\ln L$ where $L$ is the
system size. This is because the smallest $J_i$ on a chain is of
order $1/L$ and serves as a lower cutoff for the average $\langle
1/J_i \rangle$. Thus in finite but large chains the stiffness
appears to jump at the point $\a=0$, in analogy with KT transition
in a pure system. Disorder suppresses this jump logarithmically
with increasing size. Note that in the superfluid regime,
$\alpha>0$, the collective modes can be described by a harmonic
chain with off diagonal disorder (see e.g.
Ref.~[\onlinecite{alexander}]). In particular, all finite energy
phonons in the disordered superfluid are localized~\cite{matsuda}.

In this letter we studied a disordered 1-d O(2) quantum rotor
model. Using RSRG analysis, we found a strong-randomness fixed
point that controls a transition between an incompressible Mott
glass and a superfluid phase. The new phase transition is
consistent with KT universality class and has a dynamical exponent
$z=1$. Surprisingly, this fixed point lies on the classical axis
corresponding to vanishing charging energy and uniform
distribution of Josephson couplings. Unlike other fixed points
found using RSRG, this one does not belong to the infinite
randomness universality class. Nevertheless, we manage to find the
physical properties of the model in the critical region and in the
insulating and superfluid phases. One can also use the (universal)
solutions of the RSRG flow we found to calculate additional
thermodnamic quantities, as well as finite temperature properties
of the system. This we leave for future work.

{\em Acknowledgments.} We are grateful to  L. Balents, E. Demler,
D.S. Fisher, M.P.A. Fisher, D. Greenbaum, A. Paramekanti, N.
Prokof'ev, S. Sachdev and D-W. Wang for useful discussions and
comments. This work was supported by ARO (E.A.) and NSF under
grants DMR-0233773 (A.P.), DMR-0231631, DMR-0213805 (A.P., Y.K.),
DMR-0229243 (Y.K.) and PHY99-07949 (G.R.).


\begin{thebibliography}{99}
\bibitem{note} We do not imply that the distributions of $\z$ and $\b$ are
independent of each other. Rather there is a joint distribution
function $F(\z,\b,\G)$, and $f(\z,\G)=\int d\b F(\z,\b,\G)$,
$g(\b,\G)=\int d\z F(\z,\b,\G)$.
\end{thebibliography}
\end{document}